\documentclass[12pt]{iopart}
\usepackage{graphicx}

\def\bc{\begin{center}}
\def\ec{\end{center}}

\def\beq{\begin{equation}}
\def\eeq{\end{equation}}

\begin{document}

\title{Electron-phonon interaction for adiabatic anharmonic phonons}

\author{K Ziegler and D Schneider}
\address{Institut f\"ur Physik, Universit\"at Ausgburg, D-86135 Augsburg, 
Germany}
\ead{Klaus.Ziegler@Physik.Uni-Augsburg.de}

\begin{abstract}
A model with Holstein-like electron-phonon coupling is studied in the limit of
adiabatic phonons. The phonon distribution is anharmonic with two degenerate
maxima. This model can be related to fermions in a correlated binary alloy
and describes microscopic phase separation.
We discuss the weak and strong electron-phonon coupling limit
and present a qualitative phase diagram. In terms of the phononic displacements
it consists of a homogeneous, an alternating, and a disordered phase. There
is a first order phase transition between the homogeneous and the alternating
phase, and second order phase transition between the alternating and the 
disordered phase. The opening of a gap inside the disordered phase is treated 
by a dynamical mean-field theory.
\end{abstract}



\section{Introduction}

Adiabatic (or static) phonons have been discussed in the literature by
many authors \cite{holstein,millis,freericks96,freericks03}. They can 
be considered as a first step towards a full
treatment of the electron-phonon interaction in a many-body system.
The latter is plagued by a number of difficulties like the unrestricted
number of phonons at each lattice site. These problems are partially avoided
by the adiabatic limit, where the phonons are described by classical 
degrees of freedom.

In general, the electron-phonon interaction can be successfully treated
in a self-consistent perturbation theory, as long as the corresponding
coupling is small. This approach is known as the Migdal-Eliashberg theory
\cite{migdal,eliashberg}. At strong coupling, however, this theory experiences 
problems known as the breakdown of the Migdal-Eliashberg theory. In the case 
of the Holstein model it was argued by Benedetti and Zeyher that this breakdown
at large electron-phonon coupling is associated with the appearence of
an effective double-well potential for the atomic positions 
\cite{zeyher98}. It is conceivable
that the appearence of a degeneracy in the atomic positions is a
source of problems in a perturbative approach like that of Migdal and
Eliashberg.

The aim of this work is to discuss the physics of anharmonic adiabatic
phonons with two degenerate maxima of the distribution
at each lattice site in terms of a simple model.
We ignore the electronic spin which is of no interest in our study
and concentrate only on the scattering of spinless fermions. 
Our discussion includes the derivation of an effective Ising-spin model,
where the spins represent the degenerate maxima of the distribution, 
and the application of the dynamical mean-field
theory, which corresponds with the infinite-dimensional limit of the model.
Our anharmonic model can also be considered as the strong-coupling regime of
the adiabatic Holstein model.

The paper is organized as follows: The model is discussed in section 2 and an
effective distribution of the adiabatic phonons is derived in section 3. In
the limit of discrete phonon degrees of freedom (correlated binary alloy)
this is related to a distribution of Ising spins (section 3.1). The latter
is studied in the weak- (section 4.1) and strong-coupling (section 4.2) limits.
In section 5 we apply the dynamical mean-field approach to our model.
Finally, the results of the two approaches are discussed in section 6.

\section{The model for adiabatic phonons}

The Holstein-like model with spinless fermions coupled to adiabatic anharmonic
phonons is defined by the Hamiltonian $H_{\rm f}$ and the thermal phonon
distribution $P_0$ at inverse temperature $\beta$:
\beq
\fl H_{\rm f} = -{\bar t}\sum_{\langle r,r'\rangle}c_r^\dagger c_{r'}
- \sum_r\Big(\mu -g x_{r}\Big)c_{r}^\dagger c_{r}\ ,\qquad
P_0(x_r)\propto\exp\left[-\beta U(x_r^2-1)^2\right].
\label{hamilton}
\eeq
$c_{r}^\dagger$ ($c_{r}$) are the fermionic creation (annihilation)
operators and the real variable $x_r$ represents the phonon degrees
of freedom. The latter describes the displacements of the atoms
at sites $r$, assuming that it is thermally distributed according
to $P_0$. 
The fermions feel the displaced atom as a one-body potential
$-gx_r$. For harmonic phonons we have a Gaussian distribution
\beq
P_h(x_r)\propto \exp\left(-\beta x_r^2\right),
\label{harmon}
\eeq
instead of $P_0$, with $x_r=0$ as the position with maximal weight.
The anharmonic distribution $P_0$, on the other hand, has
two degenerate positions with maximal weight, namely $x_r=\pm 1$. The
coupling to the fermions, however, can break this
degeneracy, leading to homogeneous and inhomogeneous equilibrium
distributions of the atomic positions on the lattice.

Although the model defined in equation (\ref{hamilton}) looks like a
model for Anderson localization this is not a correct interpretation. 
The reason is that a
model for Anderson localization requires the averaging of the Green's
function with respect to $P_0$. This is not the case for the adiabatic 
phonons:
A grand-canonical ensemble of spinless fermions, coupled to adiabatic phonons,
is defined by the partition function
\beq
Z={\rm Tr}\; \rme^{-\beta H}
=\int{\rm Tr}_{\rm f}\; \rme^{-\beta H_{\rm f}}\prod_rP_0(x_r)\rmd x_r
\label{part1}
\eeq
which represents an {\it annealed} average of the fermionic system.
The fermionic Green's function then reads in the temperature formalism
\begin{equation}
\fl G_{r,t;r',0}= {1\over Z}{\rm Tr}\left[
\rme^{-(\beta-t) H}c_r\rme^{-tH}c^\dagger_{r'}
\right]={1\over Z}\int{\rm Tr}_{\rm f}\left[
\rme^{-(\beta-t) H_{\rm f}}c_r\rme^{-tH_{\rm f}}c^\dagger_{r'}
\right]\prod_rP_0(x_r)\rmd x_r.
\label{green1}
\end{equation}

The anharmonicity is related to previous studies where an effective
anharmonic (double-well) potential for the electrons was found 
\cite{zeyher98}. Here the main idea is that the tunneling of the fermions
is much faster than the motion of the atoms between the two maxima of
the distribution.
Therefore, the dynamics of the phonons is negligible and adiabatic
phonons can serve as phononic degrees of freedom.

\section{Effective Phonon Distribution}

The trace with respect to the fermions can be performed in
equations (\ref{part1}) and (\ref{green1}), since
the fermions do not interact directly in the Hamiltonian $H_{\rm f}$
\cite{negele}. It gives the determinant of a $N\times N$
matrix, averaged with respect to the distribution $P_0$, such that the
partition function reads \cite{ziegler02}
\[
Z=\int {\rm det}
\left[{\bf 1}+\rme^{\beta(\mu +{\hat t}- gx)}\right]\prod_r P_0(x_r)\rmd x_r.
\]
$N$ is the number of lattice sites and ${\hat t}$ is the hopping
matrix. The integrand of $Z$ is a positive expression
and can be considered as an effective distribution of the adiabatic
phonons that replaces $P_0$:
\beq
P(\{ x_r\})={1\over Z}
{\rm det}\left[{\bf 1}+\rme^{\beta(\mu +{\hat t}- gx)}\right]
\prod_rP_0(x_r).
\label{distr1}
\eeq

\subsection{Ising-Spin Representation}

For the evaluation of the effective phonon distribution given in equation
(\ref{distr1})
we consider the limiting case $U\gg1$. Then the distribution $P_0$ selects
\[
x_r\to S_r=\pm1 ,
\]
where $S_r$ is formally an Ising spin which describes discrete displacements of
the atoms. This case can be considered as a correlated binary alloy (CBA),
where the correlations are
mediated by the fermions through the determinant in equation (\ref{distr1}). The
partititon function is now a sum with respect to Ising spins:
\[
Z=\sum_{\{S_r=\pm1\}}Z(\{S_r\})\quad {\rm with} \quad
Z(\{S_r\})={\rm det}\left[{\bf 1}+e^{\beta(\mu+{\hat t}-gS)}\right].
\]
Then the fermionic Green's function is a resolvent, averaged with respect
to the effective phonon distribution \cite{ziegler02}:
\[
\left\langle \left(\rmi\omega+\mu+{\hat t}-gS\right)^{-1}
\right\rangle_{\rm Ising},
\]
where $\langle ... \rangle_{\rm Ising}$ is the CBA distribution
\beq
P(\{ S_r\})
={
{\rm det}\left[{\bf 1}+e^{\beta(\mu+{\hat t}-gS)}\right]
\over\sum_{\{S_r\}}
{\rm det}\left[{\bf 1}+e^{\beta(\mu+{\hat t}-gS)}\right]}.
\label{distra}
\eeq
It should be noticed that the CBA distribution is not $Z_2$ invariant,
i.e. not invariant under a global flip of the spins or the coupling
constant $g$, in contrast to the distribution $P_0$.

\section{Approximations of the CBA Distribution}

The effective phonon distribution given in equation (\ref{distra})
can describe different types of order of the displacements of the
atoms, depending on the inverse temperature $\beta$. At high
temperatures we expect a paramagnetic (disordered) distribution,
and at low temperatures some kind of order. Some insight can be
obtained by evaluating the distribution in the asymptotic
regimes of weak as well as strong coupling. More details of the
calculations are given in Appendix A. For further simplification
of the results
we restrict our interest to the low-temperature regime, i.e. to 
$\beta\sim\infty$.

\subsection{Weak-Coupling Limit}

The weak-coupling limit is obtained from an expansion for small
values of $g$.
According to Appendix A the regime describes an uncorrelated binary 
alloy with the distribution
\beq
P_w(\{ S_r\})=\prod_r {\rme^{-\beta ghS_r}
\over \sum_{S_r=\pm1}\rme^{-\beta ghS_r}
},
\label{weakc}
\quad{\rm with}\quad
h=\int\Theta\left(\mu+\epsilon(k)\right){\rmd^{d}k\over (2\pi)^d},
\eeq
where $\epsilon(k)$ is the dispersion of the tunneling term ${\hat t}$.
Thus the fermions create a homogeneous magnetic field $gh$ for the
Ising spins with
$0\le h\le 1$. This favors the atomic position $x_r=-1$:
\[
\langle S\rangle =-\tanh \left(\beta gh\right).
\]
For a low fermion density (i.e. for $\mu+\epsilon(k)<0$) there
is no magnetic field ($h=0$). Then the Ising spins are
paramagnetic (disordered without correlations).

\subsection{Strong-Coupling Limit}

The strong-coupling regime enables us to apply an expansion
in terms of the fermion hopping. As shown in Appendix A we can
write for the effective phonon distribution
\[
P_s\propto \prod_r\left[1+\rme^{\beta(\mu-gS_r)}\right]
\exp\Big\{
\sum_{r,r'}\left[E_1(S_r+S_{r'})+E_2S_rS_{r'}\right]
\Big\}
\]
with coefficients
\[
\fl E_1={\hat t}_{rr'}{\hat t}_{r'r}{\beta^2\over 8}\left[
{\rme^{\beta\mu-\beta g}\over(1+\rme^{\beta\mu-\beta g})^2}
-{\rme^{\beta\mu+\beta g}\over(1+\rme^{\beta\mu+\beta g})^2}
\right]\quad{\rm and}
\]
\[
\fl E_2={\hat t}_{rr'}{\hat t}_{r'r}
\left\{
{\beta^2\over 8}\left[
{\rme^{\beta\mu-\beta g}\over(1+\rme^{\beta\mu-\beta g})^2}
+{\rme^{\beta\mu+\beta g}\over(1+\rme^{\beta\mu+\beta g})^2}
\right]
-{\beta\over 4g}{\rme^{\beta\mu}\sinh(\beta g)\over
1+\rme^{2\beta\mu}+2\rme^{\beta\mu}\cosh(\beta g)}
\right\}.
\]
This gives in the low-temperature regime ($\beta\sim\infty$):
\[
E_1\sim0,\ \ \ E_{2}\sim \cases{
-{\beta\over 4g}{\hat t}_{rr'}{\hat t}_{r'r} & for $-g<\mu<g$ \cr
0 & otherwise \cr
}
\]
and for the distribution density
\beq
\fl P_s\sim\cases{
1 & for $\mu<-g$ \cr
\exp\Big\{-\beta\Big[\sum_r(\mu+g)S_r/2+{\bar t}^2
\sum_{\langle r,r'\rangle}S_rS_{r'}/4g\Big]\Big\}
 & for $-g<\mu<g$ \cr
\exp\Big(-\beta g\sum_rS_r\Big) & for $g<\mu$ .\cr
}
\label{strongc}
\eeq
In terms of the Ising spins there is a paramagnetic phase for $\mu<-g$.
For $-g<\mu<g$ the competition of the magnetic field term and the
antiferromagnetic spin-spin interaction leads to a first order
phase transition between ferro- and antiferromagnetic states, at least for
low temperatures.
A simple mean-field approximation reveals that the Ising groundstate is
an antiferromagnet for $\mu<d{\bar t}^2/g-g$ and a
ferromagnet for $\mu>d{\bar t}^2/g-g$. Finally, for $\mu>g$ there is always
a ferromagnetic state with $\langle S\rangle<0$. These three phases are
shown in the low-temperature $\mu-g$ phase diagram in figure \ref{phase}.

\section{The DMFT equations}

The spectral density of the itinerant fermions can be evaluated
conveniently by the Dynamical Mean-Field Theory (DMFT), originally
developed for the Hubbard model \cite{freericks03,metzner,georges}.
The main idea of the DMFT is the infinite-dimensional limit of the
system. This limit
is characterized by a simplified scattering process in comparison with
the situation in a finite-dimensional system: the scattering of the
{\it same} pair of fermions is very unlikely due to the large phase
space. In the case of our adiabatic phonon system it is very unlikely
that a fermion scatters with the {\it same} phonon more than once.
This enables us to consider only the effective scattering on a single
site. A consequence is that the effective correlation between the
phonons, discussed in the weak- and strong-coupling regime, is excluded
in the DMFT \cite{rem1}.

Formally, the Green's function reads
\beq
G_n=\int{\rho(\epsilon)\over \rmi\omega_n+\mu-\Sigma_n-\epsilon}\rmd\epsilon
\label{int2}
\eeq
with the Matsubara frequency $\omega_n$ and the density of states $\rho$
of free fermions on the lattice.
Moreover, the Green's function can also be written as \cite{freericks03}
\beq
G_n=\frac{w_0}{G_{n}^{-1}+\Sigma_n+g}+\frac{1-w_0}{G_{n}^{-1}+\Sigma_n-g}.
\label{green}
\eeq
Equation (\ref{green}) yields the self-consistent equation of the self energy $\Sigma_n$
:
\beq
\Sigma_n=-{1\over 2G_n}\pm{1\over2}\sqrt{G_n^{-2}+4g^2+4g(1-2w_0)G_n^{-1}},
\label{sigma}
\eeq
where the coefficient $w_0$ depends on the self energy and $G_n$ as
\[
w_0=\left(
1+\rme^{-\beta g}\prod_{n=-\infty}^\infty{\Sigma_n+G_n^{-1}+g
\over\Sigma_n+G_n^{-1}-g}
\right)^{-1}.
\]
From these three equations the self energy $\Sigma_n$ and the Green's
function $G_n$ can be determined numerically \cite{freericks03}.

\begin{figure}
\begin{center}
\includegraphics[scale=1]{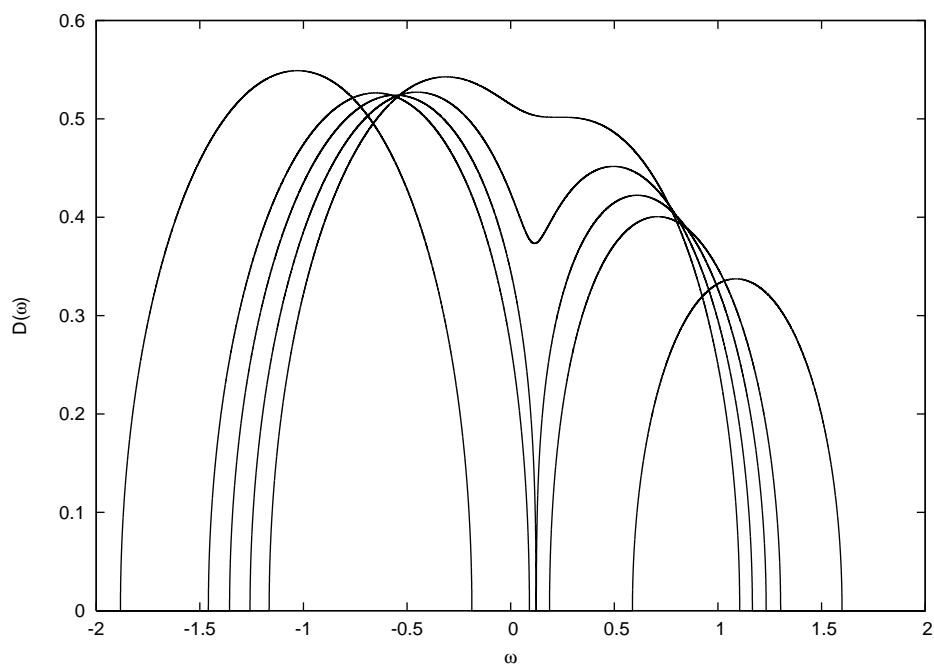}
\end{center}
\caption{\label{DOS}Density of states for varying coupling stengths 
($g=0.3$, $0.4$,
$0.5$, $0.6$ and $1.0$) for $T=1$, $\mu=0$ and ${\bar t}=0.5$. Increasing
electron-phonon-coupling favors the gap formation.}
\end{figure}

For a special form of the density of states $\rho$ the integral in
equation (\ref{int2}) can be evaluated explicitly. This simplifies the
numerical effort for solving the equations substantially.
An example is the case of a semi-circular density of states
$\rho(\epsilon)=\sqrt{4{\bar t}^2-\epsilon}/2\pi {\bar t}^2$ which gives
\cite{kotliar}
\beq
\label{green2}
G_n=\frac{1}{2{\bar t}^2}\left(\zeta_n-{\rm sgn}({\rm Re}
\zeta_n)\sqrt{\zeta_n^2-4{\bar t}^2}\right).
\eeq
with $\zeta_n=\rmi\omega_n+\mu-\Sigma_n$. $w_0$ is determined numerically using
equations (\ref{sigma}) and (\ref{green2}).
Substituting $\Sigma_n=\rmi\omega_n+\mu-G_{n}^{-1}-{\bar t}^2 G_n$ in equation
(\ref{green}) yields a cubic equation for the Green's function 
\cite{vanDongen}:
\beq
{\bar t}^4 G_{n}^3 - 2 \eta {\bar t}^2 G_{n}^2 + (\eta^2 +{\bar t}^2 -g^2)G_n -[\eta+g(1-2w_0)]=0
\eeq
with $\eta=\rmi\omega_n +\mu$. Analytic continuation ($\rmi\omega_n \rightarrow
\omega +\rmi 0^{+}$, $G_n \rightarrow G(\omega)$) and the condition Im $G(\omega)$ $<0$
leads to the fermionic density of states $D(\omega)=-{\rm Im\;}G(\omega)/\pi$.
Our results are shown in figure \ref{DOS}.

\section{Discussion of the Results}

\begin{figure}
\begin{center}
\includegraphics[scale=0.7]{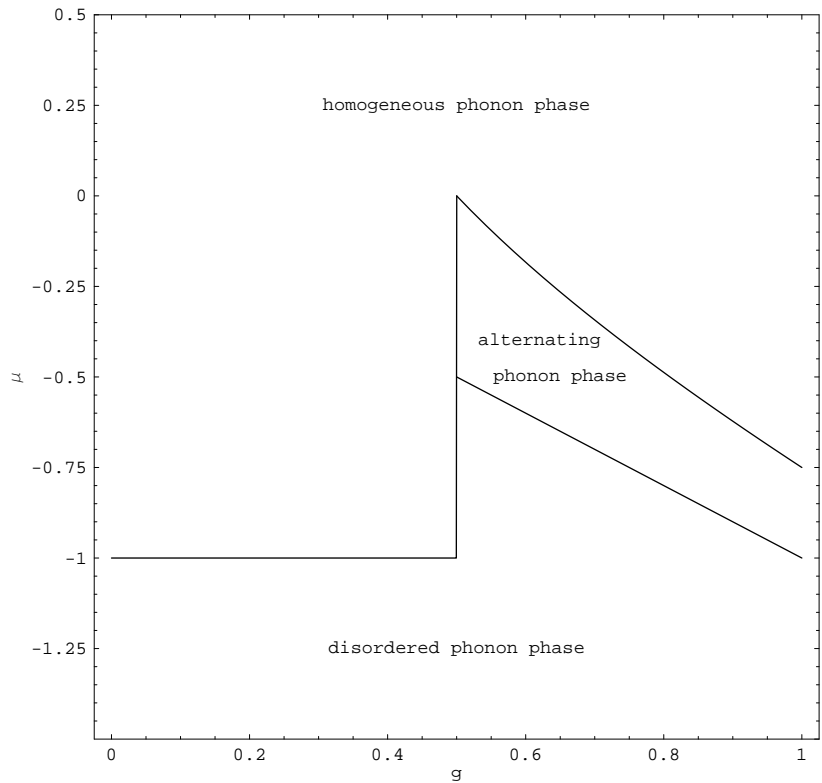}
\end{center}
\caption{\label{phase}Mean-field $T=0$--phase diagram. Thermal fluctuations 
are likely to create a disordered phase for small values of $g$ and any 
value of $\mu$.}
\end{figure}

The phase diagram of our model is complex and has been studied by
two different mean-field approaches, a classical one for the Ising-spin
representation (i.e. the CBA)
and the DMFT. It consists of para-, ferro- and antiferromagnetic
phases in terms of the Ising spins (cf. equation (\ref{strongc})). The
latter correspond to displacement configurations of the adiabatic phonons:
disordered phonons (or phonon
liquid), homogeneous and alternating (charge-ordered) phases. There are phase
transitions between these phases. According to our mean-field calculations,
a first-order transition appears between the metallic homogeneous and
the insulating alternating phase. The transition between the
disordered and insulating alternating phase is second order. The disordered
phase is most likely also insulating due to Anderson localization of the
fermions. The weak-coupling behavior of equation (\ref{weakc}) indicates a
disordered phase for $\mu<-2d{\bar t}$ and a homogeneous phase for 
$\mu>-2d{\bar t}$.
These results are summarized in the qualitative phase diagram of 
figure \ref{phase}.

\begin{figure}
\begin{center}
\includegraphics[scale=1]{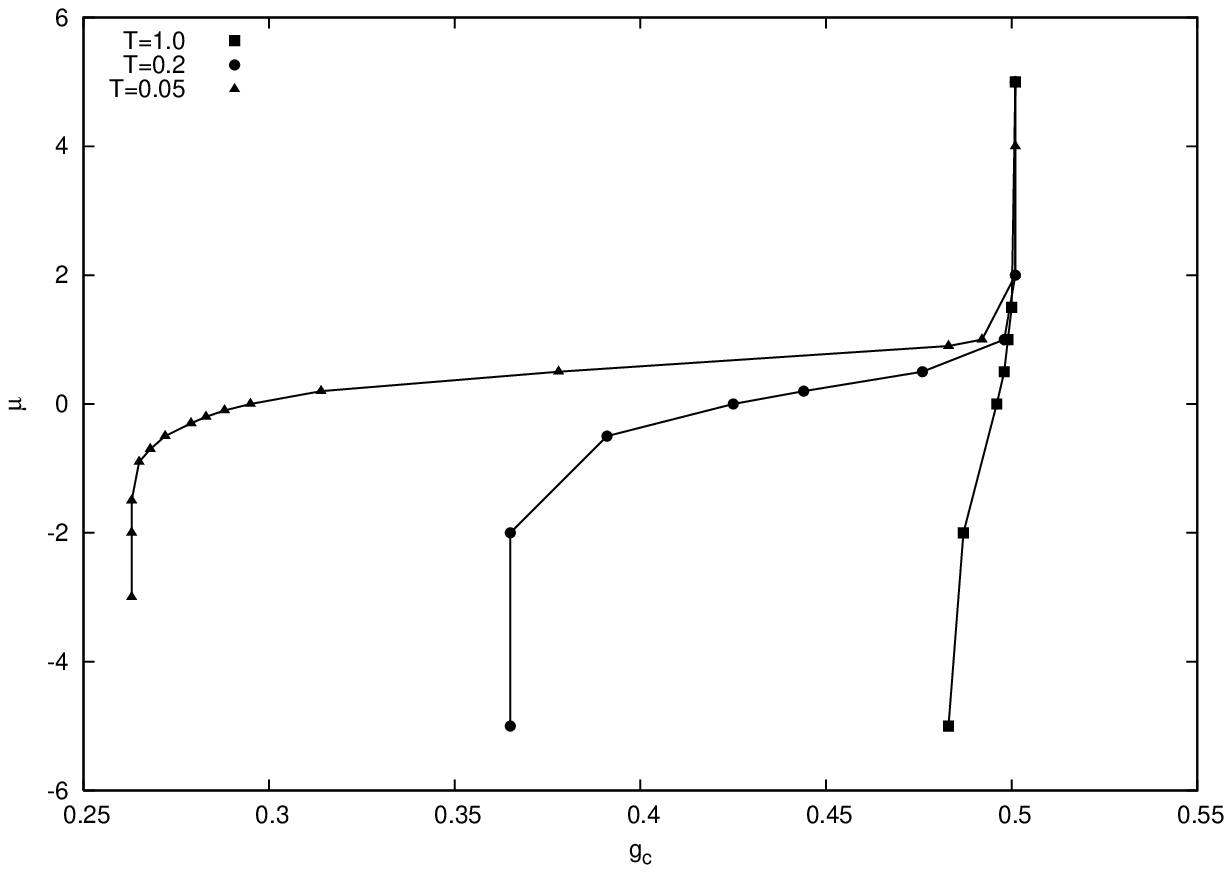}
\end{center}
\caption{\label{DMFTphase}Chemical potential versus the critical value of the electron-phonon
coupling $g_c$ plotted for ${\bar t}=0.5$ and serveral temperatures $T=0.05$, $0.2$ and
$1.0$ (from left to right). The lines are guides to the eyes.}
\end{figure}

Our DMFT results do not agree very well with the picture obtained from the
CBA. This might be a consequence of the fact that the
DMFT is a local approach, at least in the version discussed in this
paper \cite{rem1}. There is a critical coupling $g_c$ that indicates
the opening of a gap for $g>g_c$.
In our model $g_c$ changes with $\mu$ and $T$ (cf. figure \ref{DMFTphase}). 
Large positive
values for $\mu$ imply $g_c\approx 0.5$ for hopping rate ${\bar t}=0.5$.
Decreasing $\mu$ results in
a reduced value of $g_c$. The lower the temperature the lower the value of
$g_c$. Lower temperatures favour the insulating phase.

The gap opening at $g_c$ cannot be directly linked with a characteristic
change of the ordering of the adiabatic phonons, 
at least not in the regime of strong coupling and/or high temperatures.
The anharmonicity plays a crucial role here because it can open a gap
of width $2g$ already in the absence of tunneling. This can be compared
with the situation of harmonic phonons: in equation (\ref{hamilton}) 
we replace the distribution $P_0$ by the harmonic distribution in
(\ref{harmon})
and study the strong-coupling regime. The treatment of the determinant
of equation (\ref{distr1}) in strong coupling does not depend on the type of 
the random variable $x_r$. Thus we obtain the distribution for $x_r$
as given in equation (\ref{strongc}), except for an extra factor $P_h(x_r)$. 
In contrast to the distribution of the Ising spins, this distribution 
has only disordered or homogeneously ordered phases for $x_r$ but no
staggered order, implying the absence of a gap.
It remains an open
question whether or not the gap opening and the order of the adiabatic 
anharmonic phonons are related in general. This can be compared with the 
metal-insulator transition of the Hubbard model inside a paramagnetic phase
\cite{metzner,georges}. It might be that the gap opening in both cases
does not require any additional change of order in the models.
A common feature of these models is the appearence of two eigenstates which 
are separated by the interaction energy if electronic tunneling is absent.
This is fundamental also for the gap formation in the system with
electronic tunneling.
Such a separation of electronic state does not exist for adiabatic
{\it harmonic} phonons.
A remaining
puzzle is that the strong-coupling result in equation (\ref{strongc}) is an
antiferromagnetic Ising model that provides a metallic homogeneous or a gapped
alternating phase, whereas the DMFT result indicates only a gap in the
disordered phase.

Experimental relevance of our results can be linked to the complex
structures observed in manganites and related materials 
\cite{dagotto01,littlewood03}. It is believed that the interesting
properties of these materials (like the colossal magnetoresistance, or CMR)
are caused by a frozen mixture of ferromagnetic metallic and 
antiferromagnetic (or charge-ordered) insulating clusters. The 
typical size of these clusters is of the order of ten to hundred
nanometers. This has been understood as electronic phase separation
due to the proximity to a first order phase transition. Although
our model contains much less degrees of freedom than a realistic
description of manganites, the phenomenon of 
self-organized phase separation due to frozen metallic (i.e. homogeneously
ordered) and insulating (i.e. staggered) clusters and a first order
phase transition between the homogeneously ordered and the staggered
phase is also essential in our theory in terms of the correlated
binary alloy.


\section{Conclusions}

Anharmonic phonons were studied in the adiabatic limit of the Holstein model.
We applied two different approaches to study the effect of the electron-phonon
coupling, one is based on a correlated binary alloy (represented by Ising
spins),
and the other one on the dynamical mean-field theory. These approaches cover
different regions of the phase diagram. In terms of the phonons the correlated
binary alloy reveals a homogeneous, an alternating (or charge-density wave), and
a disordered phase. In terms of the fermions there is a metallic state in
the homogeneous phonon phase, and insulating states in the alternating phonon
phase (with a gap) and in the disordered phonon phase (without gap).
The dynamical mean-field theory, on the other hand, indicates a gapped
insulating state inside the disordered phase.

\ack{This work was supported by the Deutsche Forschungsgemeinschaft through
Sonderforschungsbereich 484.}

\appendix
\section{Effective Ising Models}

The determinant in equation (\ref{distra})
\beq
P={1\over Z}
{\rm det}\left[{\bf 1}+\rme^{\beta(\mu+{\hat t}-\Delta gS)}\right]
\label{distr3}
\eeq
is expanded in powers of $g$ (weak-coupling regime)
or in powers of the tunneling rate ${\bar t}$ (strong-coupling
regime).

In the weak-coupling regime we can expand the argument of the
exponential function in powers of $g$. This gives in leading order
\[
P\approx \exp\left\{
-\beta g \Tr\left[G_0\rme^{\beta(\mu+{\hat t})}S
\right]\right\}/N_w
=\exp\Big\{
-\beta g \sum_r\left[G_0 \rme^{\beta(\mu+{\hat t})}\right]_{rr}S_r
\Big\}/N_w
\]
with
$G_0=({\bf 1}+\rme^{\beta(\mu+{\hat t})})^{-1}$ and the normalization
$N_w$. $[G_0 \rme^{\beta(\mu+{\hat t})}]_{rr}$ can be evaluated by
Fourier transformation (${\hat t}\to\epsilon$) as
\[
\left[G_0 \rme^{\beta(\mu+{\hat t})}\right]_{rr}
=\int \left[1+\rme^{-\beta(\mu+\epsilon(k))}\right]^{-1}{\rmd^{d}k\over (2\pi)^d}.
\]
In the limit $\beta\to\infty$ the integrand becomes the Heavyside
step function $\Theta(\mu+\epsilon(k))$.

The expansion of the expression in equation (\ref{distr3}) in powers of
${\hat t}$ can be applied in the strong-coupling regime:
\[
\fl \exp\left\{\Tr\left[
\ln\left({\bf 1}+\rme^{\beta(\mu+{\hat t}-gS)}\right)
\right]\right\}
\approx
\det \left(G_1^{-1}\right)\exp\left[
\Tr(G_1 D)-{1\over2}\Tr(G_1 D G_1 D)
\right]
\]
\[
\fl {\rm with} \quad G_1^{-1}={\bf 1}+\rme^{\beta(\mu-gS)} \ \ \ {\rm and}\ \
D=\rme^{\beta(\mu{\hat t}+gS)} -\rme^{\beta(\mu-gS)}.
\]
A lenghty but straightforward calculation gives with
$A_r=\mu-gS_r$ the relation
\[
\Tr(G_1 D)-{1\over2}\Tr(G_1 D G_1 D)\approx{\beta\over 2}
\sum_{r,r'}
{\rme^{\beta A_r}-\rme^{\beta A_{r'}}\over A_r-A_{r'}}
{{\hat t}_{rr'}{\hat t}_{r'r}\over(1+\rme^{\beta A_r})
(1+\rme^{\beta A_{r'}})}
\]
Since there are only values $S_r=-1,1$, the term
\[
E(S_r,S_{r'})={\rme^{\beta A_r}-\rme^{\beta A_{r'}}\over A_r-A_{r'}}
{{\hat t}_{rr'}{\hat t}_{r'r}\over(1+\rme^{\beta A_r})
(1+\rme^{\beta A_{r'})}}
\]
can also be expressed as a quadratic form with respect to the
Ising spins:
\[
E(S_r,S_{r'})=E_0+E_1(S_r+S_{r'})+E_2S_rS_{r'}
\]
with
\begin{eqnarray}
\fl E_0={1\over4}\left[E(1,1)+E(-1,-1)+2E(1,-1)\right]\nonumber\\
\lo={\hat t}_{rr'}{\hat t}_{r'r}
\left\{
{\beta^2\over 8}\left[
{\rme^{\beta\mu-\beta g}\over(1+\rme^{\beta\mu-\beta g})^2}
+{\rme^{\beta\mu+\beta g}\over(1+\rme^{\beta\mu+\beta g})^2}
\right]
+{\beta\over 4g}{\rme^{\beta\mu}\sinh(\beta g)\over
1+\rme^{2\beta\mu}+2\rme^{\beta\mu}\cosh(\beta g)}
\right\},\nonumber
\end{eqnarray}
\begin{eqnarray}
\fl E_2={1\over4}\left[E(1,1)+E(-1,-1)-2E(1,-1)\right]\nonumber\\
\lo= {\hat t}_{rr'}{\hat t}_{r'r}
\left\{
{\beta^2\over 8}\left[
{\rme^{\beta\mu-\beta g}\over(1+\rme^{\beta\mu-\beta g})^2}
+{\rme^{\beta\mu+\beta g}\over(1+\rme^{\beta\mu+\beta g})^2}
\right]
-{\beta\over 4g}{\rme^{\beta\mu}\sinh(\beta g)\over
1+\rme^{2\beta\mu}+2\rme^{\beta\mu}\cosh(\beta g)}
\right\},\nonumber
\end{eqnarray}
and
\[
\fl E_1={1\over4}[E(1,1)-E(-1,-1)]
={\hat t}_{rr'}{\hat t}_{r'r}{\beta^2\over 8}\left[
{\rme^{\beta\mu-\beta g}\over(1+\rme^{\beta\mu-\beta g})^2}
-{\rme^{\beta\mu+\beta g}\over(1+\rme^{\beta\mu+\beta g})^2}
\right].
\]

\end{document}